\documentclass[12pt]{article}
\usepackage{epsfig}

\begin{document}
\title{\bf Taub, Rindler, \\ kaj la gravito de statika plato} 
\author{   A.F.F. Teixeira \thanks{teixeira@cbpf.br} \\
         {\small Centro Brasileiro de Pesquisas F\'{\i}sicas } \\ 
         {\small 22290-180 Rio de Janeiro-RJ, Brazilo} } 
\date{$3^{\rm a}$ de Februaro de 2005}
\maketitle 

\abstract
{
Estu nefinhava 3D-a plato de unuforma nepremebla fluido, kun finhava diko, 
kune kun nefinhava 2D-a plano de unuforma maso meze la plato. 
Ekvacioj de Einstein estas ekzakte solvataj, ene la 3D-a maso.  
La solvo estas kunigata en amba/u flankoj al la malena metriko de vakuo de Taub.  
/Ciu valoro por la 2D-a maso, pozitiva a/u negativa, permesas perfektan kunigon.  
Anka/u la kunigo al la malena metriko de vakuo de Rindler estas montrata;  
se neperfekta kunigo estas akceptata, denove /ciu valoro de la 2D-a maso estas ebla.   
Kelke da niaj rezultoj kontra/udiras asertojn findatajn en la literaturo.  
}
\abstract 
{
Let be an infinite 3D plate of homogeneous incompressible fluid, with finite thickness, 
together with a 2D infinite homogeneous mass in the centre of the plate. 
Einstein equations are exactly solved, in the interior of the 3D mass. 
The solution is joined on both sides to the exterior vacuum metric of Taub. 
Every value for the 2D mass, positive or negative, allows a perfect junction. 
Also the joining to a vacuum metric of Rindler is shown; 
if an imperfect joining is allowed , then again every value of the 2D mass is possible. 
Some of our results contradict assertions found in the literature. 
The text is also available in English by e-mail, ask the author. 
}

 
\newpage 
\section{Anta\u uparolo} 
\^Generala relativeco estis preparata por simili\^gi al la Newton-a gravito, kiam la gravita kampo estas malforta kaj la velo de korpoj estas malgranda.  
Ekzemple, la linigata solvo de statika vakuo kun sfera simetrio trovata de Einstein, kaj la  ekzakta solvo de Schwarzschild, amba\u u plenumas tiujn postulojn:  
amba\u u produktas kampon de akcelo kiu kunrespondas al la Newton-a (Kepler-a) $Gm/r^2$. 
Anka\u u la cilindra metriko de Levi-Civita produktas graviton simila al la Newton-a, kies kampo de akcelo estas $2G\lambda/r$.

Surprizo okazis kiam ni studis la graviton de statika vakuo kun plana simetrio.   
La Newton-a gravito anta\u udiras unuforman kampon de akcelo, kun konstanta intenso $g=2\pi G\sigma$.   
Tamen la \^generala relativeco \^sajnas anta\u udiri du malsamajn eblojn.   

La unua eblo estus la metriko de Taub, kies Riemann-a tensoro estas nenula. 
Tiu metriko anta\u udiras neunuforman kampon de akcelo, kiu ordinare forti\^gas kun la distanco al la fonto.  
La dua eblo estus la {\it flata} spacotempo de Rindler, kies Riemann-a tensoro estas nula.    
Anka\u u tiu metriko montras akcelon kiu ordinare forti\^gas kun la distanco al la fonto.  
Kelkaj a\u utoroj \cite{Synge} diras ke $R^\mu\,_{\nu\rho\sigma}=0$ signifas neekziston de gravito.  
Tamen la pliparto de la a\u utoroj \cite{Desloge2} pensas ke la postulo de ekvivalento de Einstein donas graviton al la metriko de Rindler. 
(Ja, teknike la metriko de Rindler estas aparta okazo de la Taub-a, \^car oni povas iri de Taub al Rindler per \^san\^goj de koordinatoj.) 

Do ni domandas: kiu, la Taub-a a\u u la Rindler-a, estas la korekta Einstein-a kunresponda de la Newton-a plana gravito?  
Por prepari respondon, oni ser\^cis fontojn kiuj donus amba\u u metrikojn.  
Frue, 2D-aj fontoj estis studataj (ebenoj, a\u u tre maldikaj platoj), kun finhava surfaca denso de maso, surfaca premo, kaj surfaca stre\^co \cite{2D}.  

Poste, 3D-aj fontoj (platoj kun finhava diko) estis studataj, de statikaj fluidoj. 
La ena metriko devus kuni\^gi al la Taub-a a\u u al la Rindler-a malena metriko de vakuo \cite{3D}.   

Ni havas du procedojn por fari la kunigon. 
Je la unua, \^ciuj koeficientoj de la metriko, kaj \^giaj unuaj derivoj, devas esti kontinuaj tra la surfaco de disigo.  
Je la dua, nur la koeficientoj devas esti kontinuaj.   

En \^ci tiu artikolo, ni studas du gravitajn fontojn, unu 2D-an kaj la alia 3D-an, samtempe, kaj ni havigas la ekzaktan solvon por la ena metriko.  
(La 2D-a fonto estas facile nuligata, se ni volas.)  
Poste ni kunigas tiun solvon al solvoj de vakuo. 
Unue ni kunigas al la metriko de Taub, kun \^ciuj $g_{\mu\nu}$ kaj $\partial g_{\mu\nu}/\partial x^\rho$ kontinuaj;   
poste ni kunigas al la metriko de Rindler, uzante la du eblaj procedoj por la kunigo. 

\newpage 
\section{Newton-a studo de la sistemo} 
Estu unuforma surfaca denso de maso $\sigma_0$, kiu okupas la tuta plano $z=0$.
Anka\u u estu plato kun diko $2h$, de nepremebla fluido kun unuforma 3D-a denso $\rho$=konst, okupante la tuta spaco $-\infty\!<$\{$x,y$\}$<\!\infty, \hskip2mm -h\!<\!z\!<\!h$. 
La tri parametroj $\{\sigma_0,\rho, h\}$ estas sendependaj.  
La simetrio de \^ci tiu fizika sistemo permesas skribi \^ciujn funkciojn kiel de $z>0$ nur.  
La esprimoj por $z<0$ estas poste facile havataj.  
Ni konsideras $\rho>0$, sed $\sigma_0$ povas esti $>,=,<0$. 

Ene la plato, la potencialo $\phi$ kaj la premo $p$ plenumas  
\begin{equation}                                                             \label{e01}
\phi_{int}\!''(z)=4\pi G\rho, \hskip3mm p_{Nw}\!'(z)=-\rho\phi_{int}\!'(z), \hskip3mm 0\!<\!z\!<\!h, 
\end{equation}
kie $'$ signifas $d/dz$; kaj en la malena vakuo ni havas   
\begin{equation}                                                             \label{e02}
\phi_{ext}\!''(z)=0, \hskip3mm p_{Nw}(z)=0, \hskip3mm z\!\geq h. 
\end{equation} 
Ni facile integras tiujn ekvaciojn.  
La premo, la potencialo, kaj la unua derivo de la potencialo devas esti kontinuaj tra la surfaco $z=h$,  
\begin{equation}                                                             \label{e03} 
p_{Nw}(h)=0, 
\end{equation}
\begin{equation}                                                             \label{e04}
\phi_{int}(h)=\phi_{ext}(h)=0, 
\end{equation} 
\begin{equation}                                                             \label{e05}  
\phi_{int}\,'(h)=\phi_{ext}\,'(h).  
\end{equation}

Ene la plato ($|z|\!\!<\!h$) la Newton-a potencialo, la Newton-a premo de la fluido, kaj la kampo de akceloj (turnata al la centro de la plato kiam $\sigma_0+2\rho z>0$) do estas      
\begin{equation}                                                             \label{e06}    
\phi_{int}(z)=-2\pi G\left(\sigma_0(h-z)+\rho(h^2-z^2)\right), \hskip3mm 0\!<\!z\!<\!h,  
\end{equation} \vskip-4mm 
\begin{equation}                                                             \label{e07} 
p_{Nw}(z)=2\pi G\rho\left(\sigma_0(h-z)+\rho(h^2-z^2)\right), 
\end{equation} \vskip-3mm 
\begin{equation}                                                             \label{e08}
g_{int}(z)= 2\pi G(\sigma_0+2\rho z).
\end{equation} 

Malene la plato ($|z|\!\geq\!h$), la Newton-a potencialo kaj la unuforma kampo de akceloj $g$ (turnata al la plato kiam $\sigma_0+2\rho h>0$) estas     
\begin{equation}                                                             \label{e09}             
\phi_{ext}(z)=g(z-h), \hskip3mm z\geq h, 
\end{equation} \vskip-4mm 
\begin{equation}                                                             \label{e10} 
g: = 2\pi G\sigma, \hskip3mm {\rm kie}  
\end{equation} \vskip-3mm 
\begin{equation}									           \label{e11} 
\sigma:=\frac{1}{2\pi G}\phi'(h)=\sigma_0+2\rho h  
\end{equation}  
estas la tuta surfaca masdenso de la sistemo.  
Kiam $\sigma_0=-2\rho h$, la gravito el la pozitiva kaj el la negativa maso nuli\^gas   ($\sigma=0$), kaj malena gravito neekzistas: $g=0$. 

Apud la centro de la plato $z=0^{+}$ ni findas  
\begin{equation}                                                             \label{e12}
\phi_{int}(0^{+})=-2\pi Gh(\sigma_0+\rho h), \hskip3mm  \phi_{int}\,'(0^{+})=2\pi G\sigma_0, 
\end{equation}   
\begin{equation}                                                             \label{e13}
p_{Nw}(0^{+})=2\pi G\rho h(\sigma_0+\rho h), \hskip3mm p_{Nw}\,'(0^{+})=-2\pi G\rho\sigma_0. 
\end{equation} 
Atentu ke la surfaca masdenso $\sigma_0\neq0$ sur la plano $z=0$ faras nekontinuon de la derivoj $\phi_{int}\,'$ kaj  $p_{Nw}\,'$ tra la plano;                   
por $z\rightarrow0$ je la dekstra ni havas   
\begin{equation}                                                             \label{e14} 
\sigma_0:=\frac{1}{2\pi G}\,\phi_{int}\,'(0^{+})=-\frac{1}{2\pi G\rho}\,p_{Nw}\,'(0^{+}); 
\end{equation} 
tiu \^ci rezulto estos uzata en la Einstein-a traktado.  
 
\newpage 
\section{Ena Einstein-a solvo} 
\^Ciu statika metriko kun plana simetrio estas skribebla kiel 
\begin{equation}									   	  \label{e15}
ds^2={\rm e}^{2a(z)}(dx^0)^2-{\rm e}^{2b(z)}(dx^2+dy^2)-dz^2.   
\end{equation} 
Por $T^{\mu}_{\nu}=\delta^{\mu}_{\nu}(\rho c^2,-p,-p,-p)$, la Einstein-aj ekvacioj  estas    
\begin{eqnarray}                                                        \label{e16}
G_0^0&=&2b''+3b'^2=-K\rho c^2,\hskip3mm K:=8\pi G/c^4, \\               \label{e17} 
G_x^x&=&G_y^y=a''+b''+a'^2+a'b'+b'^2=K p, \\                            \label{e18}
G_z^z&=&b'^2+2a'b'=K p; 
\end{eqnarray}
kaj la identoj de Bianchi donas  
\begin{equation}										\label{e19} 
p'=-(p+\rho c^2)a'. 
\end{equation} 

Por $\rho=$konst$>0$, la (\ref{e16}) donas la generalan solvon de $b(z)$: 
\begin{equation}                                                        \label{e20}
{\rm e}^{b(z)}=\alpha\,{\rm cos}^{2/3}(\beta+\kappa z),\hskip3mm \kappa:=\sqrt{6\pi G\rho/c^2}\equiv\sqrt{3K\rho c^2/4}, 
\end{equation} 
kie $\alpha$ kaj $\beta$ estas nedimensaj konstantoj.  
Ni elektas $b(h)=0$ kaj findas $\alpha={\rm sec}^{2/3}(\beta+\kappa h)$, do   
\begin{equation}                                                        \label{e21}
{\rm e}^{b(z)}=\left(\frac{{\rm cos}(\beta+\kappa z)}{{\rm cos}(\beta+\kappa h)}\right)^{2/3}; 
\end{equation}
notoj pri $\beta$ estos balda\u u donataj.  

Ankora\u u kun $\rho=$konst, la (\ref{e19}) donas $p(z)$ kiel funkcio de $a(z)$:  
\begin{equation}                                                        \label{e22} 
p(z)=\rho c^2\left({\rm e}^{-a(z)}-1\right),    
\end{equation} 
kie ni havos $p(h)=0$ kiam ni elektos $a(h)=0$.    
 
Ni portas (\ref{e22}) kaj (\ref{e21}) al (\ref{e18}), kaj havigas difekvacion de unua ordo por ${\rm e}^{a(z)}$,   
\begin{equation}                                                        \label{e23}
2\,b'({\rm e}^{a})'+\left(b'^2+\frac{4}{3}\kappa^2\right){\rm e}^{a}=\frac{4}{3}\kappa^2, \hskip3mm {\rm kun }\hskip3mm b'(z)=-\frac{2}{3}\kappa\,{\rm tan}(\beta+\kappa z);   
\end{equation} 
la integrado de (\ref{e23}) kun $a(h)=0$ donas   
\begin{eqnarray}\nonumber                                                         
{\rm e}^{a(z)}=1 + \frac{1}{3}{\rm sin}^2(\beta+\kappa z) \, 
\{{\cal F}\!\left({\rm sin}^2(\beta+\kappa z)\right)   
\end{eqnarray} \vskip-5mm 
\begin{equation}                                                        \label{e24}   
-\frac{{\rm sin}(\beta+\kappa h)}{{\rm sin}(\beta+\kappa z)}\frac{{\rm cos}^{1/3}(\beta+\kappa h)}{{\rm cos}^{1/3}(\beta+\kappa z)} \, 
{\cal F}\!\left({\rm sin}^2(\beta+\kappa h)\right)\},
\end{equation}
kie 
\begin{equation}                                                         \label{e25} 
{\cal F}(x):=_2\!{\rm F}_1[1,\frac{2}{3};\frac{3}{2};x], 
\end{equation} 
estante $_2{\rm F}_1$ la hipergeometria serio de Gauss 
\begin{eqnarray}	                                                      \nonumber 			 
_2{\rm F}_1[a,b;c;x]:=1+\frac{a\;b}{c}\frac{x}{1!}+\frac{a(a+1)\;b(b+1)}{c(c+1)}\frac{x^2}{2!}
\end{eqnarray} 
\vskip-2mm 
\begin{equation}      							            \label{e26}
 +\frac{a(a+1)(a+2)\;b(b+1)(b+2)}{c(c+1)(c+2)}\frac{x^3}{3!}+\dots\, . 
\end{equation} 
La (\ref{e21}), (\ref{e24}) kaj (\ref{e22}) obeas la (\ref{e17}). 
La tri parametroj $\{\beta,\rho,h\}$ estas sendependaj, kaj tute karakterizas la sistemon.

Ni volas havi Einstein-ajn parametrojn $\Sigma_0, \Sigma$ kiujn kunrespondas al la Newton-naj $\sigma_0, \sigma$. 
Por malforta kampo okazas $a'(z)\approx\phi'(z)/c^2$, kiu sugestas difini, e\^c por forta kampo, 
\begin{equation}                                                        \label{f27} 
\Sigma_0(\beta,\rho,h):=\frac{3\rho}{\kappa^2}a'(0^{+}), 
\end{equation} 
\vskip-3mm 
\begin{equation}                                                        \label{f28}
\Sigma(\beta,\rho,h):=\frac{3\rho}{\kappa^2}a'(h)=-\frac{3\rho}{\kappa^2}b'(h)=\frac{\rho}{\kappa}{\rm tan}(\beta+\kappa h). 
\end{equation}
Por malforta kampo (se amba\u u $\beta$ kaj $\kappa h$ estas malgrandaj, la (\ref{e24}) donas  
$a'(0^{+})\approx\kappa(\beta-\kappa h)/3$), ni findas 
\begin{equation}                                                        \label{f29} 
\Sigma_0(\beta,\rho,h)\approx\frac{\rho}{\kappa}(\beta-\kappa h), 
\end{equation} 
\vskip-3mm 
\begin{equation}                                                        \label{f30}  
\Sigma(\beta,\rho,h)\approx\frac{\rho}{\kappa}(\beta+\kappa h)\approx\Sigma_0+2\rho h.  
\end{equation} 

La bildo 1 donas kelkajn kurbojn $\Sigma_0/\rho h$=konst kiel funkcio de $\beta$ kaj $\kappa h$, por $\beta$ kaj $\kappa h$ amba\u u malgrandaj. 
\newpage 
 
\vspace*{3mm}
\hskip-15mm\centerline{\epsfig{file=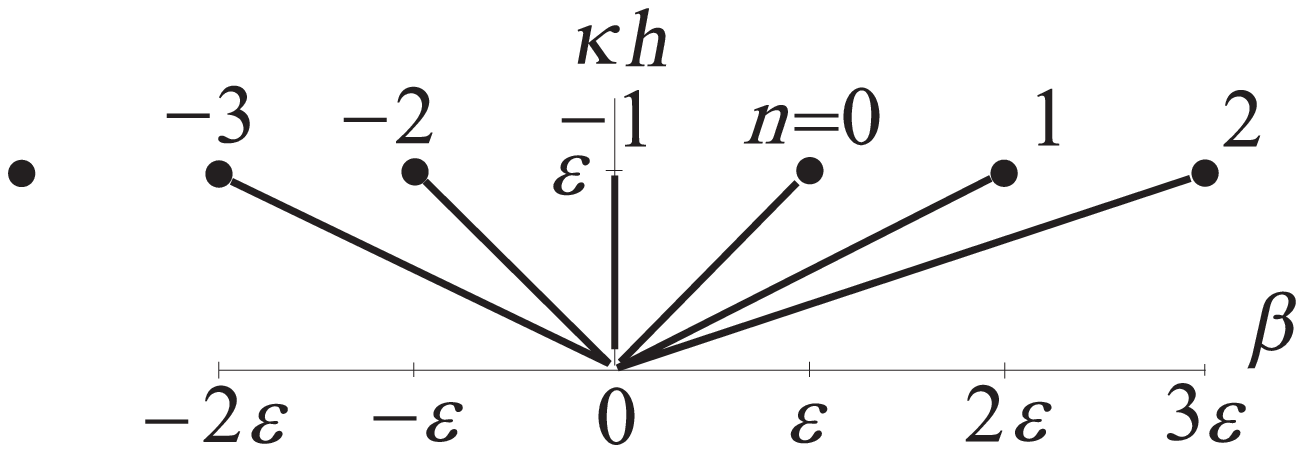,width=9cm,height=3cm}} 
\vspace*{3mm} 

\noindent {\small {\bf Bildo 1} {\it Komenco de kelkaj kurboj $\Sigma_0/\rho h=n$ kiel funkcio} (\ref{f29}) {\it de malgrandaj $\beta$ kaj $\kappa h$. 
Ni vidas ke se $\beta=\kappa h$ ni havos $\Sigma_0=0$, se $\beta=0$ ni havos $\Sigma_0=-\rho h$, kaj se $\beta=-\kappa h$ ni havos $\Sigma_0=-2\rho h$.  } } 
\vspace*{3mm} 
 
Ni povas reskribi la (\ref{f27}) kiel 
\begin{equation}                                                          \label{e30} 
\Sigma_0(\beta,\rho,h):=\frac{3\rho}{\kappa}\left(1+\frac{1}{3}{\rm tan}^2\beta-{\rm e}^{-a(0)}\right){\rm cot}\beta. 
\end{equation} 
Per komputilo, ni havigas el (\ref{e30}) la duopojn $\{\beta, \kappa h\}$ kiuj faras  $\Sigma_0=0$, prezentataj en bildo 2. 
\^Ci tiu bildo montras anka\u u la regionojn kies duopoj donas pozitivan masdenson $\Sigma_0$,   kaj $\Sigma_0$ negativan.  

\vspace*{3mm}
\centerline{\epsfig{file=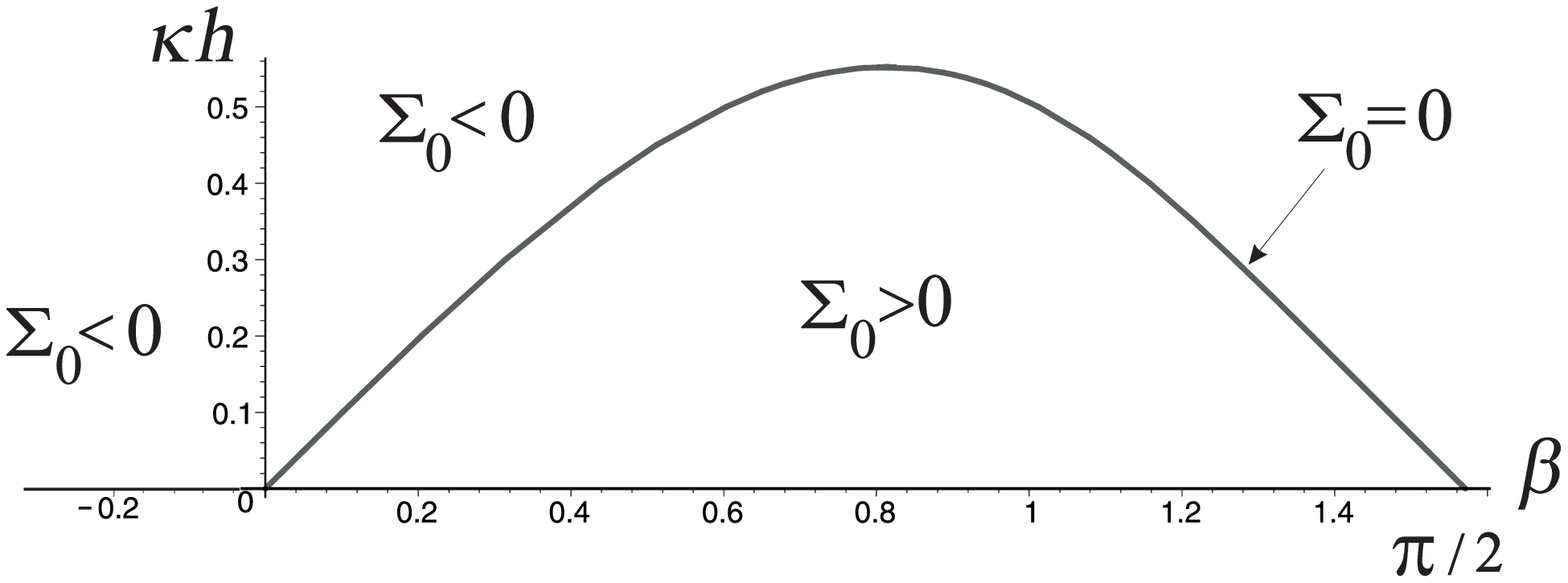,width=9cm,height=36mm}} 
\vspace*{3mm} 

\noindent {\small {\bf Bildo 2} {\it La kurbo donas la duopojn $\{\beta, \kappa h\}$ por sistemoj kun masdenso $\Sigma_0=0$. 
Sub la kurbo estas la duopoj $\{\beta, \kappa h\}$ kiuj donas pozitivan denson $\Sigma_0$. Kaj super la kurbo, inkluzive kun negativaj $\beta$, estas la duopoj $\{\beta, \kappa h\}$ kiuj donas negativan masdenson $\Sigma_0$. } } 
\vspace*{5mm} 

Poste ni kunigos la enan metrikon (\ref{e15}), (\ref{e21}), (\ref{e24}) al malenaj metrikoj, de vakuo.  
Du solvoj de vakuo ekzistas: la de Taub, kaj la de Rindler. 
\^Gi estos studataj malkune en sekcioj 4 kaj 5. 
La kunrespondaj kunados al la enaj metrikoj estos traktataj en sekcioj 6 kaj 7.   

\newpage 
\section{Vakuo de Taub}  
Ni nun studu tre simplan fizikan sistemon: la vakuan graviton el unuforma surfaca masdenso $\Sigma_0$, kiu etendi\^gas sur la plano $z=0$. 
Newton-e, la masdenso produktas potencialon (\ref{e09}), (\ref{e10}) $\phi(z)=g|z|,\hskip2mm g:=2\pi G\sigma_0$. 
Se $\sigma_0$ estas pozitiva ni havos kampon de konstanta altira akcelo $g>0$, kaj se  $\sigma_0<0$ la kampo estus anka\u u konstanta, sed pu\^sa ($g<0$). 
Ni ser\^cas metrikojn de vakuo kiuj kunrespondas al tiu sistemo.  

Por vakuo ($T^\mu_\nu=0$), la metriko (\ref{e15}) obeas  
\begin{eqnarray}                                                       \label{e31}
G_0^0&=&2b''+3b'^2=0, \\                                               \label{e32} 
G_x^x=G_y^y&=&a''+b''+a'^2+a'b'+b'^2=0, \\                             \label{e33}
G_z^z&=&(b'+2a')\,b'=0. 
\end{eqnarray}
La (\ref{e33}) havas du solvojn: $b'=-2a'$, kiu ni \^{\j}us studos, a\u u $b'=0$, kiu ni studos en sekcio 5.  

Se $b'=-2a'$ la (\ref{e31}) kaj la (\ref{e32}) estas same, kaj havas la generalan solvon ($T$ signifas Taub) 
\begin{equation}                                                        \label{e34}
{\rm e}^{a_T(z)}=\left(\frac{r}{p+qz}\right)^{1/3},\hskip3mm p,q,r={\rm konst}, 
\end{equation} \vskip-3mm 
\begin{equation}                                                        \label{e35}
{\rm e}^{b_T(z)}=(p+qz)^{2/3}. 
\end{equation} 
Por havi kampon de surfaca masdenso $\Sigma_0$ etendi\^gata sur la plano $z=0$, ni postulas $p=r=1$, $q=-3g/c^2$, kiuj donas la metrikon de Taub  
\begin{equation}                                                        \label{e36}
ds_T^2=\left(1-\frac{3g}{c^2}\,|z|\right)^{-2/3}c^2dt^2-\left(1-\frac{3g}{c^2}\,|z|\right)^{4/3}(dx^2+dy^2)-dz^2, \hskip3mm g:=2\pi G \Sigma_0.    
\end{equation}
{\it Je la} (\ref{e36}), {\it \^ciu valoro por la masdenso $\Sigma_0$ estas ebla.}  

La (\ref{e36}) havas nenulan kurbumon, \^gia skalo de Kretschmann estas 
\begin{equation}                                                        \label{e37}
{\cal K}(z):=R^{\mu\nu\rho\sigma}R_{\mu\nu\rho\sigma}=\frac{64}{27}\left(\frac{c^2}{3g}-|z|\right)^{-4}.   
\end{equation} 
Ni unue studos la okazon kun $\Sigma_0$ pozitiva ($g>0$), kiu produktas altiran gra\-vi\-tan kampon, poste ni studos la okazon kun $\Sigma_0<0$. 

Se $\Sigma_0>0$, (\ref{e37}) kun $g>0$ montras ke ${\cal K}(z)$ kreskas unuforme de ${\cal K}(0)=192g^4/c^8$ al ${\cal K}(c^2/(3g))=\infty$.                      
\^Ci tiu sugestas ke la kampo i\^gas pli forta kun la distanco al la fonto. 
Tiu malkomforto indigas studon de geodeziaj movoj, kiu ni \^{\j}us faros.  

Ni studu la movon de test-partiklo kiu, kiam $t=0$, estas senmove je $0<z_0<c^2/(3g)$.   
La geodeziaj ekvacioj donas $dx/ds=dy/ds=0$, kaj  
\begin{equation}                                                             \label{e38} 
\frac{cdt}{ds}=\frac{(1-z/z_{max})^{2/3}}{(1-z_0/z_{max})^{1/3}}, \hskip3mm \frac{dz}{ds}=-\sqrt{\left(\frac{1-z/z_{max}}{1-z_0/z_{max}}\right)^{2/3}-1}, \hskip3mm z_{max}:=\frac{c^2}{3g}. 
\end{equation}
El la (\ref{e38}) ni havigas $dz/dt$, kaj la fizikan velon de falo 
\begin{equation}                                                             \label{e39}                                                             
v_{fiz}(z):=\frac{1}{\sqrt{g_{00}}}\frac{dz}{dt}=-c\sqrt{1-\left(\frac{1-z_0/z_{max}}{1-z/z_{max}}\right)^{2/3}}  \hskip3mm (=:-c\,{\rm tanh}\xi(z)); 
\end{equation} 
vidu ke la modulo de $v_{fiz}(z)$ kreskas unuforme de 0 \^gis iu valoro $\leq c$. 
Se la partiklo ekveturas de $z_0\approx z_{max}$, \^gi falas kun $v_{fiz}\approx\,$konst$\,\approx-c$, la velo de la lumo. 

Se ni kalkulas $d^2z/dt^2$ kaj postulas $dz/dt=0$, ni havigas la akcelon de senmova korpo,   
\begin{equation}                                                             \label{e40} 
\frac{d^2z}{dt^2}=-\frac{g}{(1-z/z_{max})^{5/3}};  
\end{equation} 
vidu ke ties akcelo ne estas la fiksa Newton-a $g>0$, sed estas  $g/(1-|z|/z_{max})^{5/3}$, kiu pliforti\^gas kun $|z|$. 
Apud $z=z_{max}$, la akcelo de falo nefinhavi\^gas. 
 
Ni integras $dz/dt$ kaj findas   
\begin{equation}                                                             \label{e41}
t(z)=\frac{c}{8g}(1-z_0/z_{max})^{4/3}\left(3\xi+(\frac{3}{2}+{\rm cosh}^2\xi)\,{\rm sinh}2\xi\right), 
\end{equation}
\vskip-5mm 
\begin{equation}                                                             \label{e42}
{\rm cosh}\xi(z)=\left(\frac{1-z/z_{max}}{1-z_0/z_{max}}\right)^{1/3};
\end{equation} 
tiu \^ci estas la tempo de falo de $z_0\in(0,z_{max})$ \^gis $z\in(0,z_0)$. 
Se $gz_0<<c^2$, la (\ref{e41}) donas $z=z_0-\frac{1}{2}gt^2$, kiu estas la Newton-a rezulto.  
  
La tuta tempo de falo de test-partiklo, de $z=z_0<z_{max}$ al $z=0$ estas  
\begin{equation}                                                             \label{e43} 
t_f(z_0)=\frac{c}{8g}\left(3\alpha^4{\rm sech}^{-1}\alpha+\sqrt{1-\alpha^2}(2+3\alpha^2)\right), \hskip3mm \alpha(z_0):=(1-z_0/z_{max})^{1/3}. 
\end{equation} 
Por $gz_0<<c^2$ la (\ref{e43}) donas $t_f=\sqrt{2z_0/g}$: estas vere, en la Newton-a gravito $t_f$ kreskas unu\-for\-me kun  $z_0$. 
Sed en la Einstein-a gravito estas malsame; vidu bildojn 3, kiuj montras $t_f(z_0)$.  
Tie ni vidas kiel la pliforti\^go de la kampo kun la distanco permesas kuriozajn transpasojn.  
\newpage 

\vspace*{3mm}
\centerline{\epsfig{file=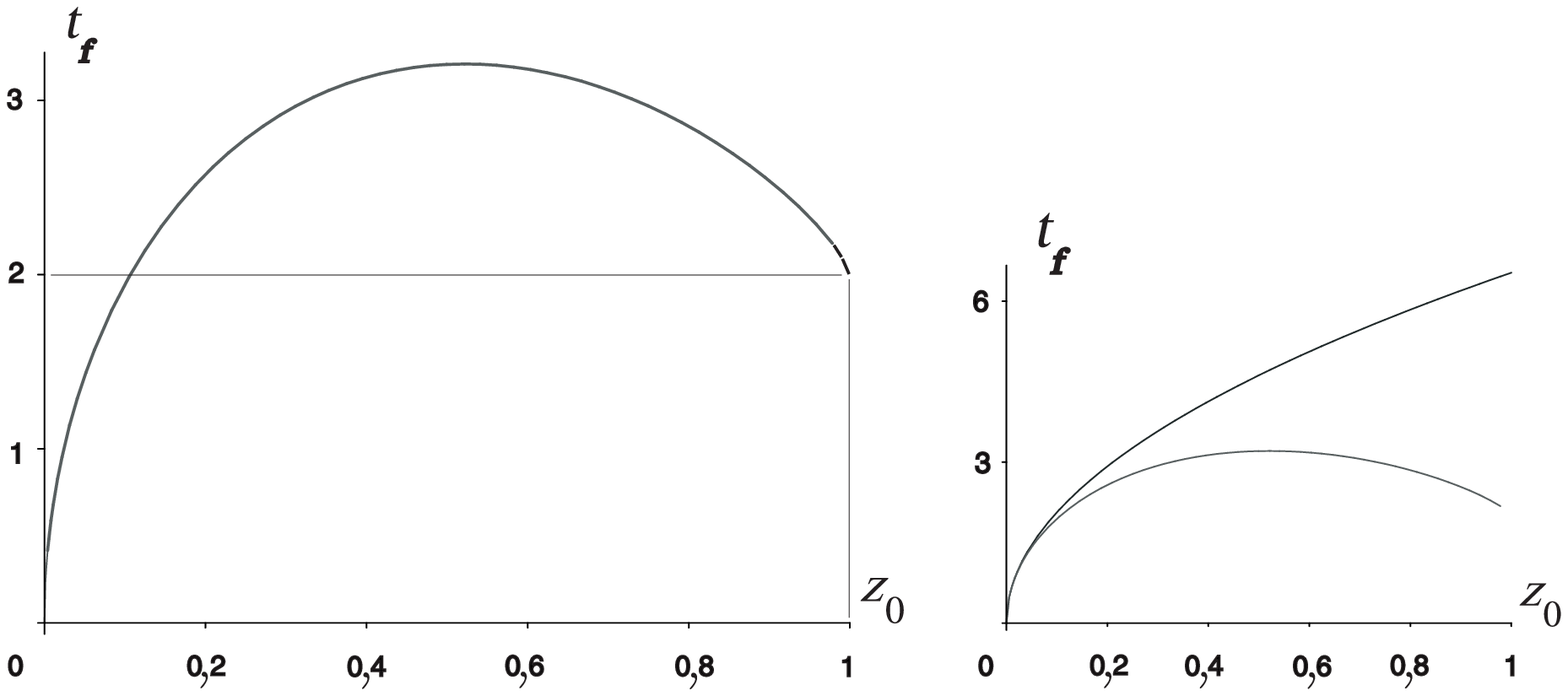,width=12cm,height=5cm}} 
\vspace*{3mm} 

\noindent {\bf Bildoj 3} {\it Pri la altira kampo de Taub: la unua bildo donas la tempon de falo  $t_f$ (ero=$c/(8g)$) de korpo kiu ekfalas, komence senmova, de $z_0$ (ero=$z_{max}$). 
Vidu ke $t_f(z_0)$ malkreskas se $z_0\!>\!0,53$. 
Sekve, korpo kiu ekfalas de $z_0>0,53$ povas transpasi, ka\u uze la pli forta komenca akcelo, aliajn korpojn kiuj ekfalas de $z_0<0,53$. 
Ekzemple, korpo kiu ekveturas de $z_0=z_{max}$ atingos $z=0$ post $t_f=2$ (same $t_f=c/(4g)$), preska\u u samtempe al alia korpo kiu ekveturas de $z_0=\frac{1}{10}z_{max}$. \\  
La dua bildo prezentas la Newton-an anta\u uvidon por la tempo de falo ($t_f=\sqrt{2z_0/g}$), kune kun la de Taub} (\ref{e43}), {\it por komparado}.  
\vspace*{5mm} 

\noindent Simpla kalkulo montras ke se $g\approx10$m/s$^2$ (tera gravito) kaj se la partiklo ekveturas apud la maksimuma distanco $|z|_{max}=c^2/(3g)\approx4$ monat-lumoj, \^gi atingos  $z=0$ post $t_f\approx3$ monatoj, kun velo $d|z|/dt\approx-c$. 
Pli generale, se $z_0\approx z_{max}$ oni havos $z_{max}\approx(4/3)ct_f$. 
Rimarku ke $t_f$ estas {\it koordinata} tempo de falo; la {\it fizika} tempo de falo, kiu estas difinata per $dt_{fiz}:=\sqrt{g_{00}}dt$, estus $\approx z_{max}/c$ se $z_0\approx z_{max}$. 

Ni nun studos la okazojn kun $\Sigma_0<0$.   
Se $g<0$ en (\ref{e37}), ni vidas ke ${\cal K}(z)$ malkreskas unuforme de ${\cal K}(0)=192g^4/c^8$  \^gis ${\cal K}(\pm\infty)=0$. 
Ni vidas, el la (\ref{e36}) kun $g<0$, ke la pu\^sa metriko de Taub validas de $z=0$ \^gis $|z|=\infty$. 
Ni domandas, kiel estas la movo de test-partiklo ekveturante kiam $t=0$, komencante senmove de $z=z_0>0$?  
La geodeziaj ekvacioj donas  
\begin{equation}                                                           \label{e44} 
\frac{cdt}{ds}=\frac{(1+3g'z/c^2)^{2/3}}{(1+3g'z_0/c^2)^{1/3}}, \hskip3mm \frac{dz}{ds}=\sqrt{\left(\frac{1+3g'z/c^2}{1+3g'z_0/c^2}\right)^{2/3}-1}, \hskip3mm g':=2\pi G|\Sigma_0|>0.
\end{equation} 
Havante $dz/dt$ ni findas, por $z=z_0$, 
\begin{equation}                                                            \label{e45} 
\frac{d^2z}{dt^2}=\frac{g'}{(1+3g'z/c^2)^{5/3}};  
\end{equation} 
ni vidas ke senmova korpo sub la Taub-a pu\^sa akcelo sentas, ne la fiksa Newton-a $g'$, sed   $g'/(1+3g'|z|/c^2)^{5/3}$, kiu malforti\^gas kun la distanco $|z|$ de la korpo al la fonto.  

Ni demandas se, kiel en la altira okazo, denove eblas transpason inter kor\-poj kiuj ekveturis samtempe, komencante senmove de nesimilaj $z$.    
Por havi respondon, ni integras $dz/dt$ kaj findas  
\begin{equation}                                                       \label{e46}
t(z)=\frac{c}{8g'}(1+3g'z_0/c^2)^{4/3}\left(3\xi+(\frac{3}{2}+{\rm cosh}^2\xi){\rm sinh}2\xi\right), 
\end{equation}
\vskip-5mm
\begin{equation}                                                       \label{e47}
{\rm cosh}\xi(z):=\left(\frac{1+3g'z/c^2}{1+3g'z_0/c^2}\right)^{1/3},                                                          
\end{equation}
kiu estas la pasanta tempo inter la pozicio $z_0>0$ kaj la $z>z_0$, de la korpo. 
Kiam ni elektas iu ajn valorojn por $g'>0$ kaj por $z>0$, kaj ni studas la kurbon (\ref{e46}) de $t$ kontra\u u $z_0\in(0,z)$, ni vidas ke denove la eblo de transpaso okazas.  
Se $g'z<<c^2$ la (\ref{e46}) donas $z=z_0+\frac{1}{2}g't^2$, kiu estas la Newton-a rezulto.  

La fizika velo de pu\^so estas  
\begin{equation}                                                           \label{e48} 
v_{fiz}(z):=\frac{1}{\sqrt{g_{00}}}\frac{dz}{dt}=c\sqrt{1-\left(\frac{1+3g'z_0/c^2}{1+3g'z/c^2}\right)^{2/3}} \hskip3mm(=c\,{\rm tanh}\xi(z)), 
\end{equation} 
kiu kreskas unuforme de 0 (kiam $z=z_0$) \^gis $c$ (kiam $z\rightarrow\infty$). 

Resume, la altira kampo de Taub ((\ref{e36}) kun $g>0$) faras la trispacon esti plato kun diko $2|z|_{max}=2c^2/(3g)$; des pli malproksime al la fonto, des pli forta estas la kampo.  
Kontra\u ue, la pu\^sa kampo de Taub ((\ref{e36}) kun $g<0$) ampleksas \^gis $|z|=\infty$;  des pli malproksime al la fonto, des pli malforta estas la kampo. 

\newpage 
\section{Vakuo de Rindler} 
Ni revenas al la ekvacioj de vakuo (\ref{e31})--(\ref{e33}). 
La okazo $b'=0$ donas $b=$konst, la (\ref{e31}) estas plenumata, kaj la (\ref{e32}) i\^gas  plisimple $a''+a'^{2}=0$, kies generala solvo estas ($R$ subskribata signifas Rindler) 
\begin{equation}                                                           \label{e49} 
{\rm e}^{a_R(z)}=P+Qz, \hskip3mm P,Q={\rm konst}, 
\end{equation} 
\begin{equation}                                                           \label{e50} 
{\rm e}^{b_R}=R={\rm konst}. 
\end{equation} 
Simile kiel en sekcio 4, ni havigas la vakuan metrikon de Rindler por surfaca unuforma masdenso  $\Sigma_0$ sur la plano $z=0$:  
\begin{equation}                                                         \label{e51}
ds_R^2=\left(1+\frac{g|z|}{c^2}\right)^2c^2dt^2-dx^2-dy^2-dz^2, \hskip3mm g:=2\pi G\Sigma_0. 
\end{equation} 

Ni unue studas la okazo $\Sigma_0$ pozitiva ($g>0$), kiu produktas altiran kampon de akcelo. 
Kontra\u ue al Taub kun $\Sigma_0>0$, la altira metriko de Rindler validas de $z=0$ \^gis  $|z|=\infty$;  
en \^ci tiu aspekto la altira kampo de Rindler estas simila al la ordinara Newton-a kampo.  
Ni konsideras test-partiklon senmovan en $z=z_0>0$ kiam $t=0$, poste \^gi libere falas la\u u la direkto al la plano $z=0$. 
La geodeziaj ekvacioj donas  
\begin{equation}                                                         \label{e52}
\frac{cdt}{ds}=\frac{1+gz_0/c^2}{(1+gz/c^2)^2}, \hskip3mm \frac{dz}{ds}=-\sqrt{1-\left(\frac{1+gz/c^2}{1+gz_0/c^2}\right)^2}. 
\end{equation}
Ni integras $dz/dt$ kaj findas  
\begin{equation}                                                         \label{e53}
z(t)=\frac{c^2}{g}\left((1+\frac{gz_0}{c^2}){\rm sech}\frac{gt}{c}-1\right). 
\end{equation}
Se $gt<<c$, ni findas $z(t)\approx z_0-\frac{1}{2}g(1+gz_0/c^2)t^2$;  
\^ci tio montras ke la altira kampo de akcelo de Rindler havas formon $g(1+g|z|/c^2)$, do \^gi kreskas linie kun $|z|$.  
Se ni faras $z=0$ en (\ref{e53}), ni havigas la tempon $t_f$ de falo
\begin{equation}                                                         \label{e54}
t_f(z_0)=\frac{2c}{g}{\rm sinh}^{-1}\sqrt{\frac{gz_0}{2c^2}}. 
\end{equation}
Vidu ke des pli granda estas la komenca distanco $z_0$, des pli granda estas $t_f$, simile al la Newton-a rezulto; sed kontra\u ue al la altira Taub, kiu permesas la transpasojn vidatajn en la bildoj 3.  
La fizika velo de falo estas  
\begin{equation}                                                         \label{e55} 
v_{fiz}(z):=\frac{1}{\sqrt{g_{00}}}\frac{dz}{dt} = -c\sqrt{1-\left(\frac{1+gz/c^2}{1+gz_0/c^2}\right)^2} \hskip3mm (=-c\,{\rm tanh}\frac{gt}{c});  
\end{equation}
vidu ke $|v_{fiz}|$ kontinue kreskas dum la falo; \^gi atingas $\approx c$ apud $z=0$ se $gz_0>>c^2$.

Ni nun studas la okazon $\Sigma_0<0$. 
La (\ref{e51}) kun $g<0$ montras ke $g_{00}$ estas nula en $|z|=z_{max}:=c^2/|g|$. 
Test-partiklo, senmova en $z_0>0$ (sed $<z_{max}$) kiam $t=0$, estas poste pu\^sata kontra\u u la direkto al la plano $z=0$, kiel   
\begin{equation}                                                             \label{e56}
z(t)=z_{max}\left(1-(1-\frac{z_0}{z_{max}}){\rm sech}\frac{g't}{c}\right), \hskip3mm g':=2\pi G|\Sigma_0|.
\end{equation} 
Vidu ke, se $g't<<c$, okazas $z(t)\approx z_0+\frac{1}{2}g'(1-z_0/z_{max})t^2$;  
\^ci tio montras ke la pu\^sa kampo de akcelo de Rindler havas formon $g'(1-|z|/z_{max})$, do \^gi malkreskas kun $|z|$. 
Ni inversigas la (\ref{e56}) kaj havigas la da\u uron de movo, de $z_0>0$ \^gis $z>z_0$, 
\begin{equation}                                                             \label{e57} 
t(z)=\frac{c}{g'}{\rm cosh}^{-1}\left(\frac{z_{max}-z_0}{z_{max}-z}\right).  
\end{equation} 
Vidu ke $t$ malkreskas unuforme kun la kresko de $z_0$, se ni fiksas \^ciujn aliajn variantojn;  
do la transpasoj kiuj okazas en la pu\^sa kampo de Taub ne okazas en la pu\^sa kampo de Rindler.  
Vidu anka\u u ke la partiklo atingas $z=z_{max}$ nur kiam $t=\infty$.  
La fizika velo de pu\^so estas  
\begin{equation}                                                             \label{e58} 
v_{fiz}(z):=\frac{1}{\sqrt{g_{00}}}\frac{dz}{dt}=c\sqrt{1-\left(\frac{z_{max}-z}{z_{max}-z_0}\right)^2} \hskip3mm (=c\,{\rm tanh}\frac{g't}{c});
\end{equation} 
vidu ke $v_{fiz}(z)$ kreskas kontinue de 0 en $z=z_0$ \^gis $\approx c$ apud $z=z_{max}$. 

Ni vidas ke la partiklo atingas finhava $z_{max}$ kun finhava velo $c$ post nefinhava tempo; 
\^ci tiu fakto estas simila al la falo de test-partiklo sur la radio de Schwarzschild.
\^Gi okazas \^car $g_{00}=(1-|z|/z_{max})^2$ etendas al nulo kiam $|z|\rightarrow z_{max}$. 

Ni komparu la Taub-an (T) al la Rindler-a (R) kampo: \\ 
$\bullet$ la altiraj kampoj, kaj T kaj R, amba\u u forti\^gas kun la distanco al la fonto; \\
$\bullet$ la pu\^saj kampoj, kaj T kaj R, amba\u u malforti\^gas kun la distanco al la fonto;\\ $\bullet$ la  pu\^sa kampo T, kaj la altira kampo R, amba\u u ampleksas la tutan trispacon $-\infty<z<\infty$; \\  
$\bullet$ la altira kampo T, kaj la pu\^sa kampo R, amba\u u ampleksas nur la platon $-z_{max}<z<z_{max}$; je T estas $z_{max}=c^2/(3g)$, kaj je R estas $z_{max}:=c^2/|g|$.
Vidu bildojn 4, kiu komparas la Newton-an potencialon al la Taub-a kaj la Rindler-a log($\sqrt{g_{00}}$). 
\newpage  
\vspace*{3mm}
\hskip-15mm\centerline{\epsfig{file=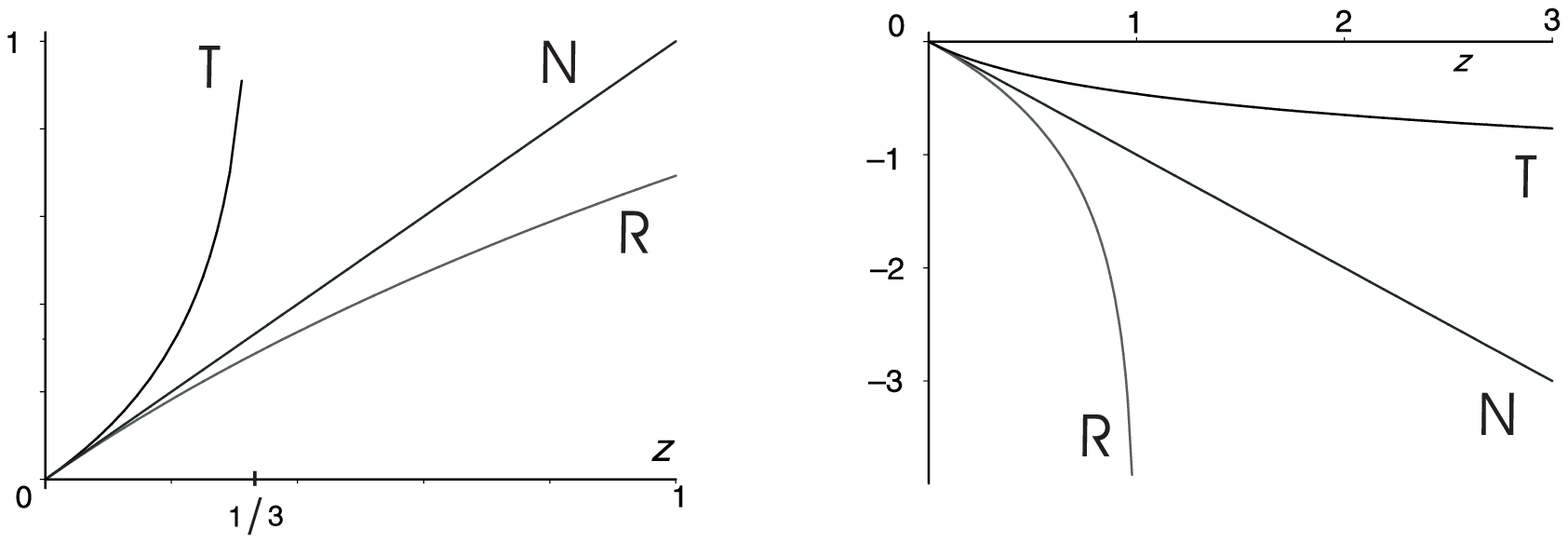,width=11cm,height=4cm}} 
\vspace*{3mm} 

\noindent  {\bf Bildoj 4} {\it Maldekstre, por $\Sigma_0>0$ $(g>0)$, estas la altiraj statikaj potencialoj de Newton $(N=z)$, de Taub $(T=-\frac{1}{3}{\rm ln}(1-3z))$, kaj de Rindler $(R={\rm ln}(1+z))$, por komparado; 
vidu ke T nefinhavi\^gas en $z=1/3$. \^Ci tie ni uzis $c=g=1$. \\ 
Dekstre, por $\Sigma_0<0$ $(g<0)$, estas la pu\^saj statikaj potencialoj de Newton $(N=-z)$, de Taub $(T=-\frac{1}{3}{\rm ln}(1+3z))$, kaj de Rindler $(R={\rm ln}(1-z))$, por komparado; 
vidu ke R nefinhavi\^gas en $z=1$. \^Ci tie ni uzis $c=-g=1$.   }  
\vspace*{3mm} 
 
Atentu ke la tensoro de Riemann de la metriko de Rindler (\ref{e51}) estas nula; do kelkaj a\u utoroj \cite{Synge} opiniis ke tiu metriko ne prezentas graviton.   
Tamen, se la kampoj estas malfortaj, ni vidis ke test-partikloj sub la metriko de Rindler veturas simile al sub la Newton-a gravito (kaj por $\Sigma_0>0$, kaj por $<0$); 
tiu \^ci faris plimulte a\u utoroj \cite{Desloge2} opinii ke $R^\mu_{\,\,\nu\rho\sigma}=0$ ne implikas mankon de gravito.  

\newpage 
\section{Plato kaj vakuo de Taub} 
Ni nun kunigas la platon kun masdenso $\rho\neq0$, kun metrikaj koeficientoj (\ref{e20}) kaj (\ref{e24}), al la ena vakuo de Taub kun (\ref{e34}) kaj (\ref{e35}). 

Ni postulas la tri kondi\^cojn de kontinuo 
\begin{eqnarray} \nonumber 
a_T(h)=a(h)\,\,(=0),
\end{eqnarray} 
\vskip-1cm 
\begin{eqnarray}                                                          \label{e59}
a'_T(h)=a'(h)\,\, (=\frac{1}{3}\kappa {\rm tan}(\beta+\kappa h)),
\end{eqnarray} 
\vskip-1cm 
\begin{eqnarray} \nonumber 
b_T(h)=b(h)\,\,(=0) 
\end{eqnarray}
kaj havigas 
\begin{eqnarray} \nonumber 
r=1,
\end{eqnarray} 
\vskip-1cm 
\begin{eqnarray}                                                         \label{e60}
 p=1+\kappa h\,{\rm tan}(\beta+\kappa h),
\end{eqnarray} 
\vskip-1cm 
\begin{eqnarray} \nonumber 
q=-\kappa\,{\rm tan}(\beta+\kappa h); 
\end{eqnarray}
la kvara kondi\^co de kontinuo, $b'_T(h)=b'(h)$, rezultas de $a'_T(h)=a'(h)$, kiel (\ref{f28}).  
Do ni findas, por $z>h$,    
\begin{equation}                                                           \label{e61}
{\rm e}^{a_T(z)}=[1-\kappa(z-h){\rm tan}(\beta+\kappa h)]^{-1/3}, 
\end{equation} \vskip-3mm 
\begin{equation}                                                           \label{e62}
{\rm e}^{b_T(z)}=[1-\kappa(z-h){\rm tan}(\beta+\kappa h)]^{2/3}, 
\end{equation} 
kiuj donas la metrikon de Taub kunatan al la sistemo $\{\beta,\rho,h\}$:  
\begin{eqnarray}      \nonumber
ds_T^2=[1\!-\!\kappa(z\!-\!h){\rm tan}(\beta\!+\!\kappa h)]^{-2/3}c^2dt^2
\end{eqnarray} 
\vskip-1cm 
\begin{equation}                                                        \label{e63}
\end{equation} 
\vskip-1cm 
\begin{eqnarray} \nonumber                                                           
-[1\!-\!\kappa(z\!-\!h){\rm tan}(\beta\!+\!\kappa h)]^{4/3}(dx^2\!\!+\!dy^2)-dz^2. 
\end{eqnarray} 

Se ni komparas la (\ref{e63}) al la (\ref{e36}) ni havigas la rilaton de la konstanto $g$ de akcelo de Taub al la tri parametroj $\{\beta,\rho,h\}$ de la plato: 
\begin{equation}                                                        \label{f63} 
g=\frac{\kappa c^2}{3}{\rm tan}(\beta+\kappa h). 
\end{equation} 
Kaj se ni skribus $g=2\pi G\Sigma$, kie $\Sigma$ estus la tuta surfaca masdenso de la sistemo, ni rehavigus $\Sigma=(\rho/\kappa)$tan$(\beta+\kappa h)$, kiel en (\ref{f28}). 

Vidu ke {\it ekzistas Taub-an solvon por iu ajn kombino} $\{\Sigma_0,\rho,h\}$. Bildoj 5 kaj 6 montras kelkajn ekzemplojn, kun $\Sigma_0>,=,<0$.  

\newpage 

\vspace*{3mm}
\centerline{\epsfig{file=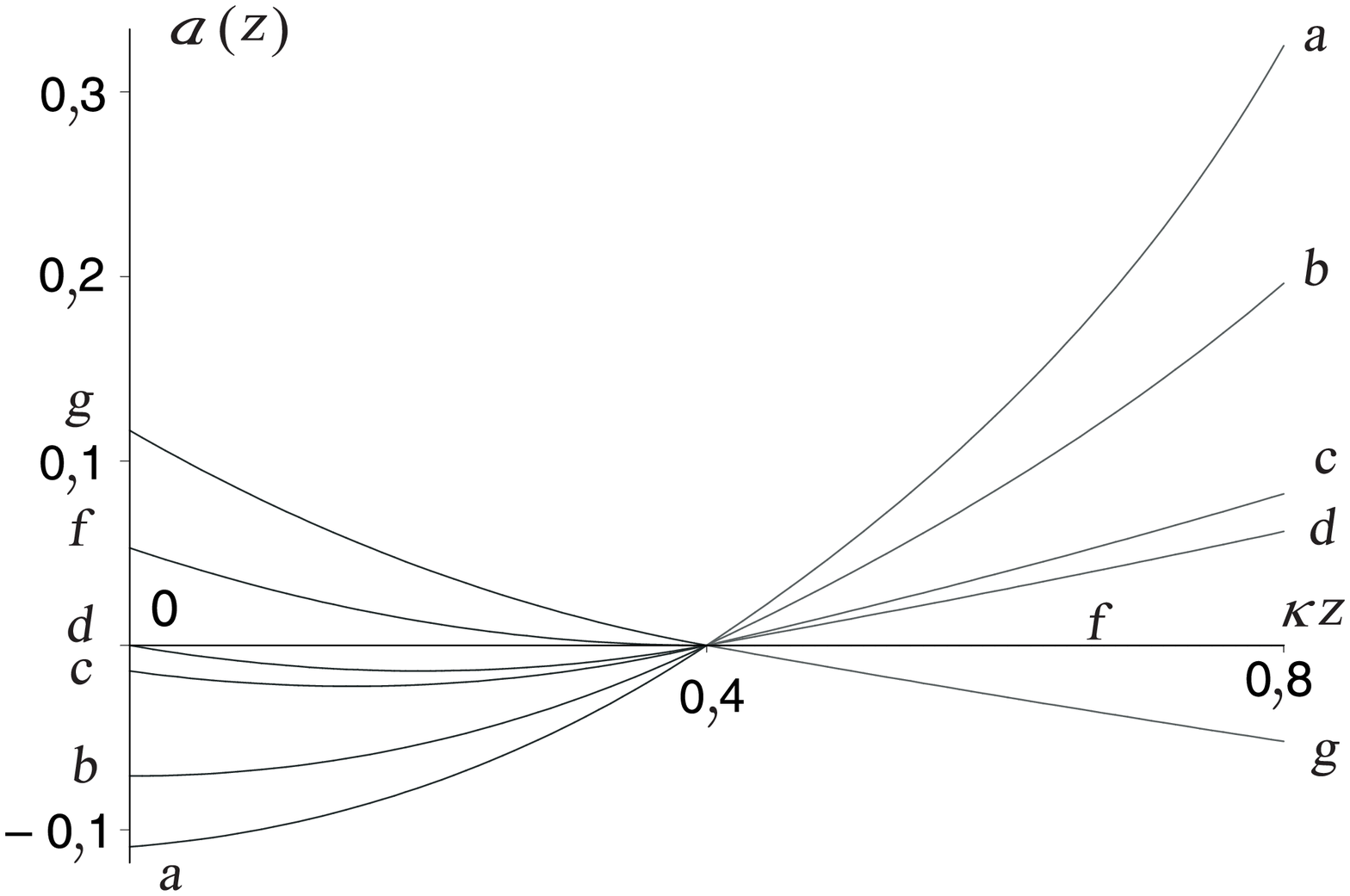,width=11cm,height=7cm}} 
\vspace*{3mm} 

\noindent {\bf Bildo 5} {\it La potencialo de la fluido} (\ref{e24}) {\it kaj de Taub} (\ref{e61}) {\it kiel funkcioj de $\kappa z$, kun $\kappa h=0,4$. 
La signo de la derivo $a'(0)$ estas la signo de la surfaca masdenso $\Sigma_0$, 
kaj la signo de la derivo $a'(h)$ estas la signo de $\beta+\kappa h$. 
\^Ciuj derivoj $a'(z)$ estas kontinuaj tra $z=h$.} 

{\bf aa} $\beta\!=\!0,6$ ; $\,\,\Sigma_0>0$, {\it akcelo \^ciam altira kaj kreskanta, nefinhavi\^gas en }  $z=h+$$\kappa^{-1}{\rm cot}(\beta+\kappa h)$; 

{\bf bb} $\beta\!\approx\!0,4387$ ; $\,\,\Sigma_0=0$, {\it akcelo komence nula, poste kiel en } {\bf aa}; 

{\bf cc} $\beta\!=\!0,1$ ; $\,\,-\rho h\!<\!\Sigma_0\!<\!0$, {\it akcelo komence pu\^sa (ka\u uze  $\Sigma_0\!<\!0$), poste kiel en } {\bf aa};  

{\bf dd} $\beta\!=\!0$ ; $\,\,-2\rho h\!<\!\Sigma_0\!<\!-\rho h$, $\,\,a(0)\!=\!0$, {\it akcelo kiel en } {\bf cc}; 

{\bf ff} $\beta\!=\!-\kappa h\!=\!-0,4$ ; $\,\,\Sigma_0\!\approx\!-2\rho h$, {\it akcelo komence pu\^sa (ka\u uze $\Sigma_0\!<\!0$), poste nula por $z\!\geq\!h$ (Lorentz);} 

{\bf gg} $\beta\!=\!-0,8$ ; $\,\,\Sigma_0\!<\!-2\rho h$, {\it akcelo \^ciam pu\^sa (ka\u uze la alta valoro de negativa $\Sigma_0$), sed malforti\^gas \^gis $z=\infty$}.

\vspace*{5mm} 

\newpage 

\vspace*{3mm}
\centerline{\epsfig{file=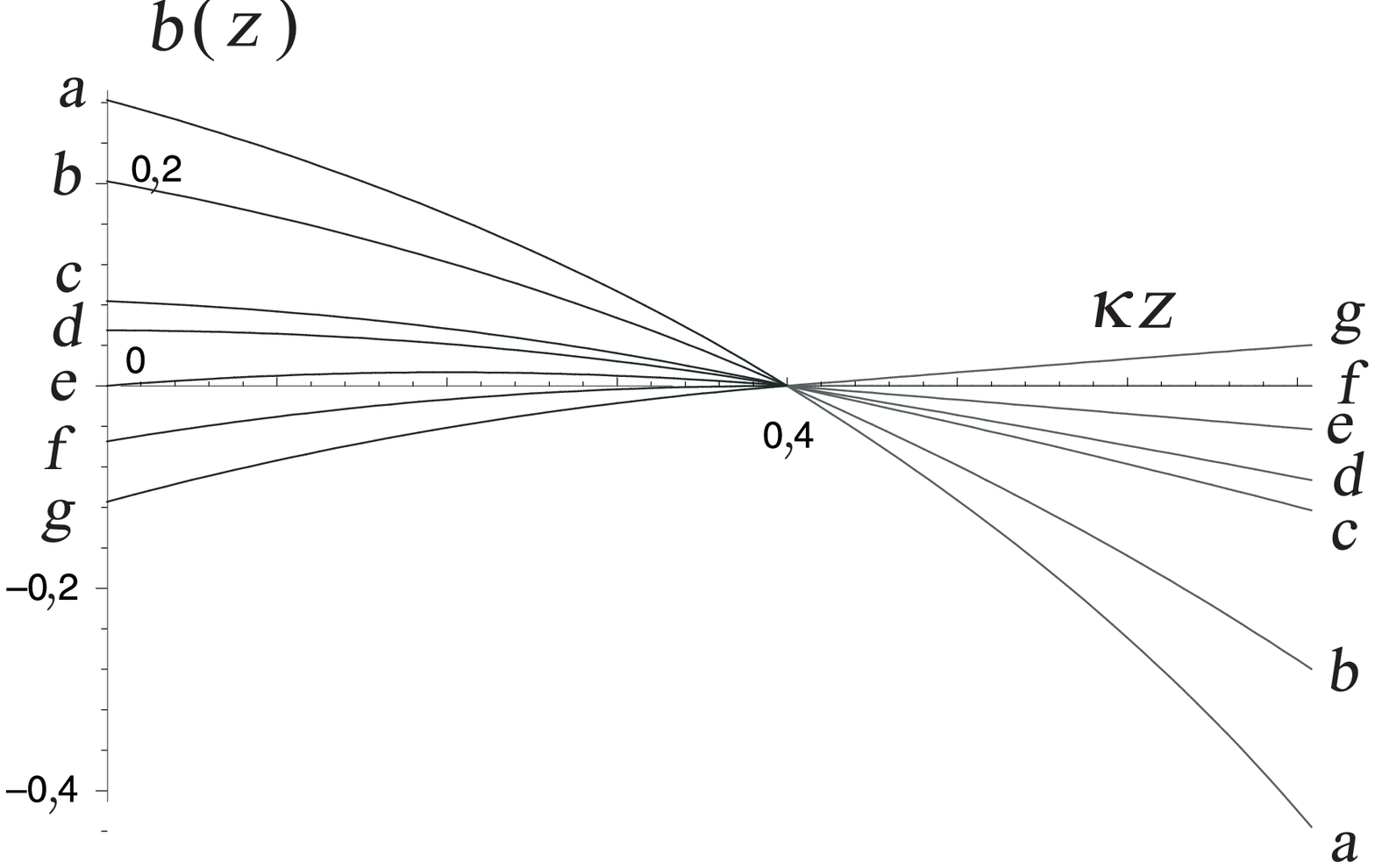,width=11cm,height=7cm}} 
\vspace*{3mm} 

\noindent  {\bf Bildo 6} {\it La potencialo} (\ref{e21}) {\it de la fluido kaj} (\ref{e62}) {\it de Taub, kiel funkcioj de $\kappa z$, por $\kappa h=0,4$. 
La signo de la derivo $b'(0)$ estas mala al de $\beta$, kaj la signo de la derivo $b'(h)$ estas mala al de $\beta+\kappa h$. 
\^Ciuj derivoj $b'(z)$ estas kontinuaj tra $z=h$.} 

{\bf aa} $(\beta=0,6), \hskip2mm$ {\bf bb} $(\beta\approx0,4387), \hskip2mm$ {\bf cc} $(\beta=0,1), \hskip2mm$ {\it kaj } {\bf dd} $(\beta=0),$ {\it \^ciuj havas $b(z)$ \^ciam malkreskantajn};   

{\bf ee} $(\beta\!=\!-\kappa h/2\!=\!-0,2)$ {\it havas $b(z)$ komence kreskantan, poste malkreskantan;} 

{\bf ff} $(\beta\!=\!-\kappa h\!=\!-0,4)$ {\it havas $b(z)$ komence kreskantan, poste nuli\^gas por $z\!\geq\!h$ (Lorentz);} 

{\bf gg} $(\beta\!=\!-0,8)$ {\it havas $b(z)$ \^ciam kreskantan.} 
\vspace*{5mm}

\newpage 
Se $\beta+\kappa h>0$ (do la (\ref{f28}) faras $\Sigma\!>\!0$), ni vidas ke la metriko de Taub (\ref{e63}) nefinhavi\^gas $(g_{00}\rightarrow\infty$, $g_{xx}\rightarrow0$) en la koordinato $z=z_{max}:=h+c^2/(6\pi G\Sigma)$. 
Vidu ke $z_{max}$ malkreskas kiam $\Sigma$ kreskas.

Se $\beta+\kappa h\approx0$ kaj $\kappa(z-h)\approx0$, la koeficiento $g_{00}(z)$ de la (\ref{e63}) konduktas al la Newton-a potencialo $\phi_{ext}(z)$ de la (\ref{e09}). 
Fakte, en tiu okazo tan$(\beta+\kappa h)\approx\beta+\kappa h\approx\kappa\Sigma/\rho$, do  
\begin{equation}                                                            \label{e64} 
[1-\kappa(z-h){\rm tan}(\beta+\kappa h)]^{-1/3}\approx 1+\frac{2\pi G\Sigma}{c^2}(z-h). 
\end{equation} 
Por malforta kampo e$^{a_T(z)}\approx1+\phi_{ext}(z)/c^2$, do la Newton-a potencialo estus  
\begin{equation}                                                            \label{e65} 
\phi_{ext}(z)\approx2\pi G\Sigma(z-h), 
\end{equation} 
kiu efekte koincidas kun (\ref{e09})--(\ref{e11}). 

La (\ref{e63}) degeneras al metriko de Lorentz (\ref{e69}) kiam $\beta+\kappa h=0$. 
Ni facile komprenas la kialon:   
kun $\Sigma=0$, la gravito nuli\^gas.   

Fine, se $-\pi/2<\beta+\kappa h<0$, la metriko de Taub (\ref{e63}) estas valebla por $h<z<-\infty$. 
 
Rimarku ke, je la kurboj $a,b,c,d$ de la bildo 5, la gradiento $a_T^{'}(z)$ estas pozitiva kaj \^ciam kreskanta; ni refindas \^ci tie la nekomfortan konduton de la altira kampo de Taub ($\Sigma\!>\!0$), kiu estas des pli forta des pli malproksime al la fonto, \^gis $z_{max}$. 

\newpage 
\section{Plato kaj vakuo de Rindler} 
Ni nun kunigas la metrikon de fluido, (\ref{e20}) kaj (\ref{e24}), al la metriko de vakuo de Rindler, (\ref{e49}) kaj (\ref{e50}).  
Ni postulas la tri kondi\^cojn de kontinuo
\begin{eqnarray} \nonumber 
a_R(h)=a(h)\,\,(=0), 
\end{eqnarray} 
\vskip-1cm 
\begin{eqnarray}                                                          \label{e66}
a'_R(h)=a'(h)\,\,(=\frac{1}{3}\kappa\,{\rm tan}(\beta+\kappa h)),
\end{eqnarray} 
\vskip-1cm 
\begin{eqnarray} \nonumber 
b_R(h)=b(h)\,\,(=0), 
\end{eqnarray}
kaj havigas  
\begin{equation}                                                           \label{e67} 
P=1-\frac{1}{3}\kappa h\,{\rm tan}(\beta+\kappa h), \hskip5mm Q=\frac{1}{3}\kappa\,{\rm tan}(\beta+\kappa h), \hskip5mm R=1;  
\end{equation} 
la metriko de Rindler kunigata al sistemo $\{\Sigma_0,\rho,h\}$ do estas, por $|z|>h$,  
\begin{equation}                                                           \label{e68}
ds_R^2=[1+\frac{1}{3}\kappa(|z|-h){\rm tan}(\beta+\kappa h)]^2c^2dt^2-dx^2-dy^2-dz^2. 
\end{equation} 
Se ni komparas la (\ref{e68}) al la (\ref{e51}), ni findas ke la Rindler-a konstanto de akcelo $g$ denove estas $g=(\kappa c^2/3)$tan$(\beta+\kappa h)=2\pi G\Sigma$, kiel en la Taub-a okazo (\ref{f63}).

Se $0\leq\beta+\kappa h<\pi/2$, do $\Sigma\!>\!0$ kaj $g\!>\!0$, la koeficiento $g_{00}$ estas fizike bona por $|z|\in(h,\infty)$.  
Sed se $\beta+\kappa h<0$, do $\Sigma\!<\!0$ kaj $g\!<\!0$, la koeficiento $g_{00}$ nuli\^gas en la plano $z=z_{max}:=h+3\kappa^{-1}{\rm cot}|\beta+\kappa h|=h+c^2/|g|$. 
Vidu ke en la limo $\{\beta+\kappa h,\kappa(z-h)\}\approx0$ la $g_{00}$ de Rindler (\ref{e68}) koincidas kun la $g_{00}$ de Taub (\ref{e63}), do denove ni havigas la potencialo  $\phi_{ext}(z)$ de Newton (\ref{e09}). 

En \^ci tiu studo de Rindler ni ne postulis ke la derivoj $g_{xx,z}$ e $g_{yy,z}$ estu kontinuaj tra la surfaco $z=h$.                                                            
Se ni postulas $b'(h)=b'_R(=0)$, ni havigas en (\ref{f28}) la kondi\^con $\beta+\kappa h=0$, do $\Sigma\!=\!0$. 
Tiuokaze la metriko de Rindler (\ref{e68}) degeneras al la de Lorentz, 
\begin{equation}                                                            \label{e69} 
ds^2=c^2dt^2-dx^2-dy^2-dz^2.  
\end{equation}
Ni ripetu, {\it la nura metriko de Rindler kiu perfekte kuni\^gas (kun $g_{\mu\nu}$ kaj $g_{\mu\nu,\rho}$ kontinuaj) al sistemo $\{\Sigma_0,\rho,h\}$ estas la metriko de Lorentz, kaj la kondi\^co $\beta+\kappa h=0$, do $\Sigma=0$, estas postulata. } 

Sistemoj $\{\Sigma_0,\rho,h\}$ kun $\beta+\kappa h=0$ estas atentintaj. 
La koeficientoj (\ref{e24}) kaj (\ref{e21}), rilataj al la plato, simpli\^gas al  
\begin{equation}                                                            \label{e70} 
{\rm e}^{a(z)}= 1+\frac{1}{3}{\rm sin}^2\kappa(h-z){\cal F}\left({\rm sin}^2\kappa(h-z)\right) \hskip2mm {\rm por} \hskip2mm\beta=-\kappa h,  
\end{equation} 
\begin{equation}                                                            \label{e71} 
{\rm e}^{b(z)}={\rm cos}^{2/3}\kappa(h-z) \hskip2mm {\rm por} \hskip2mm\beta=-\kappa h. 
\end{equation} 
Ni kalkulu $\Sigma_0$ kiam $\beta=-\kappa h$, uzante (\ref{e30}) kaj (\ref{e70}): 
\begin{equation}                                                            \label{e72} 
\Sigma_0=-\frac{3\rho}{\kappa{\rm tan}\kappa h}\left(1+\frac{1}{3}{\rm tan}^2\kappa h-[1+\frac{1}{3}{\rm sin}^2\kappa h{\cal F}({\rm sin}^2\kappa h)]^{-1}\right), \hskip1mm {\rm por} \hskip1mm \beta=-\kappa h. 
\end{equation}
Bildo 7 montras kiel $\Sigma_0/(2\rho h))$ \^san\^gas kun $\kappa h$. 
Vidu ke $\Sigma_0<0$ por \^ciu $\kappa h>0$;  
tiu montras ke anka\u u en generala relativeco nur negativaj masoj povas malfortigi la graviton de la pozitiva $\rho$. 
Speciale vidu ke, en la limo $\kappa h\rightarrow0$, okazas $\Sigma_0\rightarrow-2\rho h$; 
tiu konfirmas ke nur la platoj kun tuta masdenso $\Sigma_0+2\rho h=0$ ampleksas al la vakuo kiel Lorentz. 

\vspace*{3mm}
\centerline{\epsfig{file=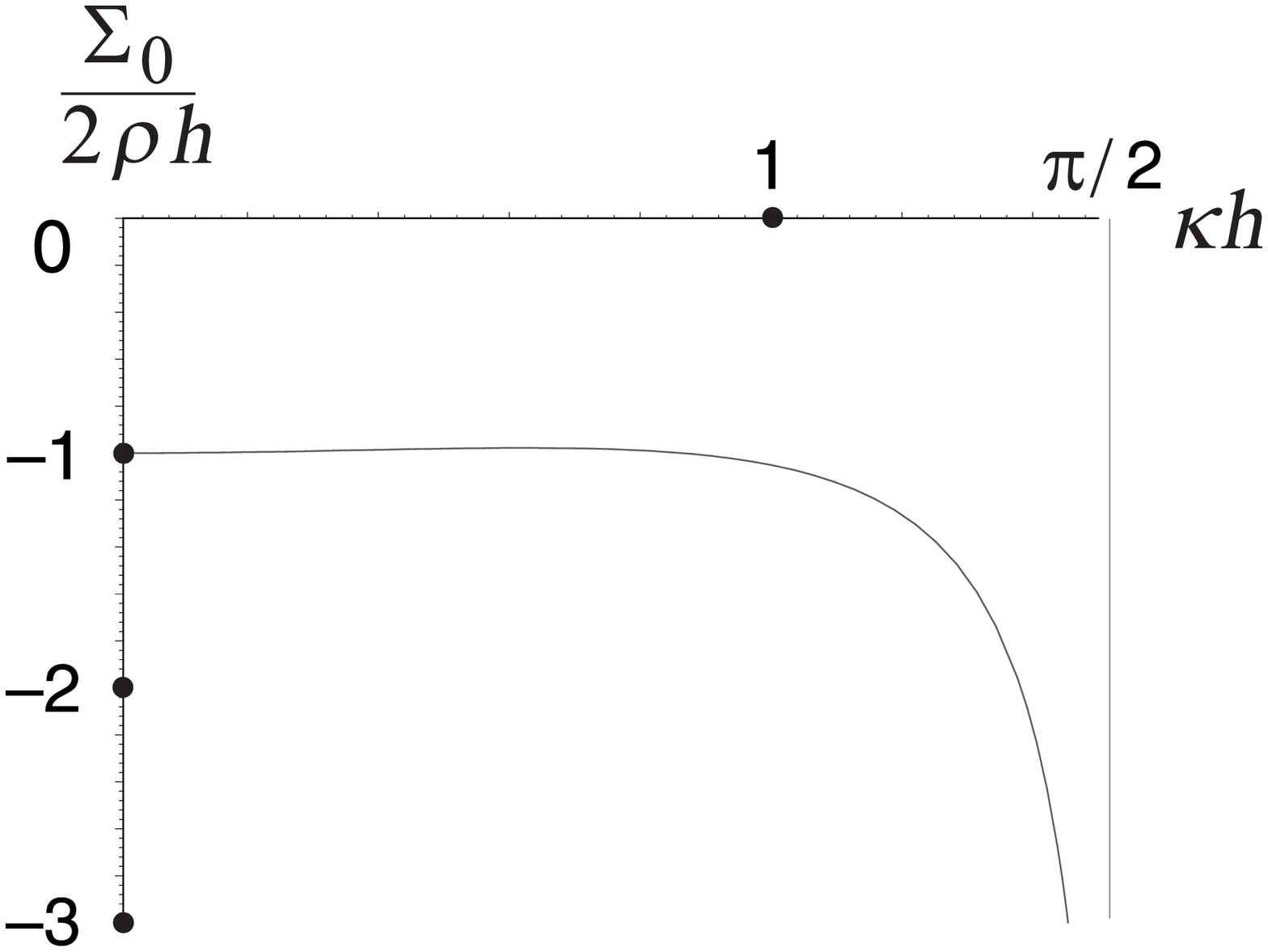,width=55mm,height=36mm}} 
\vspace*{3mm} 

\noindent  {\bf Bildo 7} {\it La kurbo estas solvo de }(\ref{e72}) {\it kaj donas la duopojn }\{$\Sigma_0/(2\rho h), \kappa h$\} {\it kiuj produktas metrikon de Lorentz} (\ref{e69}) {\it ekstere la plato. 
Por $\kappa h<1$ okazas $\Sigma_0\approx-2\rho h$, kiu estas la Newton-a kondi\^co. 
Sed por kreskanta $\kappa h$, bezonas $\Sigma_0$ esti pli kaj pli negativan, por ne\u utraligi la graviton de $2\rho h$. }  
\vspace*{5mm} 

\newpage  
\section{Komentarioj} 
La metriko de Taub estas ofte simpligata al $g_{00}=z^{-2/3}, \,\,g_{xx}=g_{yy}=-z^{4/3}$, anstata\u u sub la generala formo (\ref{e34}),\,(\ref{e35}). 
Tio simpligata formo kunrespondas al nia $r=1,p=0,q=1$, kiu kunrespondas al tuta masdenso  $\Sigma$ sufi\^ce negativa. 
Tamen, en sekcio 6, bildoj 5 kaj 6, ni pruvis ke {\it la (generala) metriko de Taub permesas \^ciu valoro de} $-\infty<\Sigma_0<\infty$. 

Ni povus anstata\u uigi $\Sigma_0$ $\leftrightarrow$ plato kun unuforma voluma masdenso $\rho_0$, kaj $2h_0$ dike. 
Tiel ni eliminus la surfacan masdenson sur la plano $z=0$, kaj okazigus $\phi'$ kaj $p'$ esti kontinuaj en la nova regiono.   
Ni kredas ke la novaj premoj kaj la novaj potencialoj estus similaj al la de \^ci tiu artikolo.   

\section{Dankoj kaj helpo} 
Ni kore dankas al kreintoj de Maple, Corel, PCTeX, eps.fig, kiu multe simpligis la kalkulojn kaj la montradon de \^ci tiu verko. 
Same al ekipoj de Microsoft kaj arxiv.org. \\
$\bullet$ Uma vers\~ao concisa deste artigo, em portugu\^es, est\'a dispo\-n\'{\i}\-vel por e-mail; pe\c ca ao autor. \\

\end{document}